\documentclass[a4paper]{article}
\usepackage{amssymb,amsmath,amsfonts}
\usepackage[dvips]{graphicx}
\newtheorem{rules}{Rule}

\usepackage{url}

\begin{document}

\title{Cryptographic requirements for chaotic secure communications}
\author{Gonzalo \'{A}lvarez\thanks{Instituto F\'{\i}sica Aplicada,
CSIC, Serrano 144, 28006, Madrid, Spain, e-mail:
gonzalo@iec.csic.es} \: and Shujun Li\thanks{Department of
Electronic Engineering, City University of Hong Kong, 83 Tat Chee
Avenue, Kowloon Toon, HK SAR, China, e-mail: hooklee@mail.com}}
\date{}

\maketitle

\begin{abstract}
In recent years, a great amount of secure communications systems
based on chaotic synchronization have been published. Most of the
proposed schemes fail to explain a number of features of
fundamental importance to all cryptosystems, such as
implementation details, or key definition, characterization, and
generation. As a consequence, the proposed ciphers are difficult
to realize in practice with a reasonable degree of security.
Likewise, they are seldom accompanied by a security analysis.
Thus, it is hard for the reader to have a hint about their
security and performance. In this work we provide a set of
guidelines that every new cryptosystem would benefit from adhering
to. The proposed guidelines address these two main gaps, i.e.,
correct key management and security analysis, among other topics,
to help new cryptosystems be presented in a more rigorous
cryptographic way. Also some recommendations are made regarding
some practical aspects of communications, such as implementation,
channel noise, limited bandwith, and attenuation.
\end{abstract}

\begin{keywords}
{Chaos; cryptography; design guidelines}
\end{keywords}

\section{Introduction}
\label{sec:intro}

Modern telecommunications networks, and specially Internet and
mobile telephones, have expanded the possibilities of user
communications and information transmission to limits unimaginable
a short time ago. There is a parallel growing demand of
cryptographic techniques, which has originated an intense research
activity and the search of new directions in cryptography. As a
result, a rich variety of chaotic cryptosystems for end to end
communications have been put forward \cite{asocscs,cc}, whose
robustness and privacy are equally diverse.

There exist two main approaches to chaotic ciphers design: analog
and digital. The first one is based on the concept of chaotic
synchronization \cite{sics}. In these systems, the information can
be transmitted by the chaotic signal in a number of ways: chaotic
masking
\cite{edoscvcs,aswtscswatscs,acmsbuscs,solbccwatc,doancbocoafi},
in which the analog message signal $m(t)$ is added to the output
of the chaotic generator $x(t)$ in the transmitter; chaotic
switching or chaos shift keying (CSK)
\cite{cskmadoaccuss,todsbcs}, in which a binary message signal is
used to choose between different chaotic attractors; chaotic
modulation \cite{ssctmocicc,cioscwatc,aacssatsc,ascsbotpsocs}, in
which a binary message modulates a parameter of the chaotic
generator or when spread spectrum techniques are used to multiply
the message signal by the chaotic one; and chaos control
\cite{cwc93,ecocfc,cwcutdsd}, in which small perturbations cause
the symbolic dynamics of a chaotic system to track a prescribed
symbol sequence. Regardless of the method used to transmit the
message signal, the receiver has to synchronize with the
transmitter's chaotic generator to regenerate the chaotic signal
$x(t)$ and thus recover the message $m(t)$.

The second approach to the design of chaos-based cryptosystems
consists of using digital computers to iterate a chaotic map and
mask the binary message in a number of ways
\cite{cwc98,scbotdcm,dccuek,amccm,natce,ctcms}. These ciphers do
not depend on synchronization. Instead, they usually use one or
more chaotic maps where the initial point $x_0$ and the parameter
value $\lambda$ play the role of the key.

Most papers on this topic are published in physics and engineering
journals and conferences, but not within the cryptography
community. This explains why up to date little or no critical
analysis has been made about the design process of these
cryptosystems nor to the way the results are presented. Quoting
Feng Bao \cite{coancac}: ``The common annoying feature of the
cryptosystems based on some mathematical models, e.g., those based
on chaos systems, is that only the principle is given. They lack
details, such as recommended key sizes and key generation steps,
etc. Therefore it is not possible for others to implement the
ciphers.''

We have detected that a systematic approach to the design and
security evaluation is missing. To fill this void, in this paper
we give some guidelines which should benefit new chaos-based
cryptosystems. Following these guidelines, proposed cryptosystems
would be presented in a more rigorous cryptographic way.
Otherwise, they tend to be information concealment methods to
frustrate the casual eavesdropper, but in no case the determined
attacker and will not be taken seriously by cryptologists.

The paper is organized as follows. Sec.~\ref{sec:systems} lists
requirements about the minimum practical details of the chaotic
implementations that should be provided. In Sec.~\ref{sec:key} the
most important key related issues are discussed. In
Sec.~\ref{sec:analysis} the recommendations about how to make a
security analysis are given. In Sec.~\ref{sec:channel} some basic
but decisive considerations about the channel properties are
offered. Sec.~\ref{sec:conclusion} concludes the paper.

\section{The Implementation}
\label{sec:systems}

For many chaotic systems, only basic concepts are described and
implementation details are neglected. However, generally speaking,
implementation details are important for cryptanalysts to judge
whether or not there exist security defects. Also, it is obvious
that the encryption speed and the implementation cost depend on
such details. Therefore, the lack of implementation details makes
it difficult to estimate the significance of the proposed
cryptosystem, i.e. to analyze its security and overall
performance.

\subsection{The implementation of the chaotic systems}

As we introduced in Sec.~\ref{sec:intro}, there are two basic
approaches to design chaotic ciphers: analog and digital. The
first one is generally based on chaos synchronization, and the
concerned chaotic systems are implemented in analog form. The
second one is independent of chaos synchronization and the chaotic
systems are completely implemented in digital computers.
Additionally, in some conditions, the chaotic systems may be
implemented in combined form, that is to say, both analog and
digital parts are involved.

When analog parts are involved, the circuit diagrams to generate
chaos should be given with enough details. When digital parts are
involved, the following details should be given: the finite
computing precision, the adopted digital arithmetic (fixed-point
or floating-point), hardware/software configuration, etc. When
both analog and digital parts are involved, details on A/D and D/A
interfaces should also be described.

\begin{rules}
It should be thoroughly described how to implement the chaotic
systems.
\end{rules}

When the chaotic systems are (completely or partially) implemented
in digital form, dynamical degradation will occur. This problem
has been extensively studied in the last two decades, and it has
been clarified that such dynamical degradation may cause security
defects in chaos-based cryptosystems. To overcome this problem,
some methods should be used to improve dynamical degradation of
digital chaotic systems. A considerable countermeasure is to
timely perturb the chaotic systems with a small pseudo-random
signal. For more details, please refer to \S2.5 of
\cite{ShujunLi:Dissertation2003}, in which Shujun Li gives a
comprehensive review on this issue.

\begin{rules}
For chaotic systems implemented in digital form, the negative
dynamical degradation should be taken into consideration.
\end{rules}

\subsection{The implementation of the cryptosystem}

In the cryptography community, there are two well-known sayings:
``it is quite easy to design a secure but very SLOW cipher'', and
``it is quite easy to design a secure but very LARGE cipher''. If
the security of a chaotic cryptosystem is reached with the loss of
working efficiency, then its significance will be trivial and will
not be accepted by pure cryptographers.

\begin{rules}
Without loss of security, the cryptosystem should be easy to
implement with acceptable cost and should work with acceptable
speed.
\end{rules}

\section{The key}
\label{sec:key}

A fundamental aspect of every cryptosystem is the key. An
algorithm is as secure as its key. No matter how strong and well
designed the algorithm might be, if the key is poorly chosen or
the key space is small enough, the cryptosystem will be broken.
Unfortunately, most chaotic secure communications schemes proposed
to date fail to clearly, if at all, explain what the key is, how
it should be chosen, and to thoroughly describe the available key
space.

\subsection{Key definition}

A cryptosystem cannot exist without a key. Otherwise, it might be
considered as a coding system, but never regarded as a secure
system. In every cryptosystem an important effort must be devoted
to clearly define and characterize the key. Many chaos-based
secure communication systems proposed in the literature do not
specify what the key is. It is assumed that the key must be made
from the parameters of the chaotic system, but it is not clearly
stated which parameters are, which their range is and what their
precision or sensibility is.

\begin{rules}
What the key is should be thoroughly defined.
\end{rules}

\subsection{The key space}

Once the key has been defined, it is equally important to
characterize it, i.e., the key space must be studied in depth.

The size of the key space is the number of encryption/decryption
key pairs that are available in the cipher system. The symbol $v$
is used to denote a key and the symbol $\mathbf{V}$ to denote a
set of keys. Since the total number of possible keys is equal to
$r=|\mathbf{V}|$, the set $\mathbf{V}$ or key space can be
expressed as
\begin{equation}\label{V}
     {\mathbf{V}}=\{v_1,v_2,\ldots,v_r\}.
\end{equation}

In classical cryptographic algorithms based on number theory, the
key is usually a string of random bits generated by some automatic
process. If the key is $n$ bits long, then every possible $n$-bit
key must be equally likely, with probability $2^{-n}$.

\begin{figure}[t]
\center \includegraphics{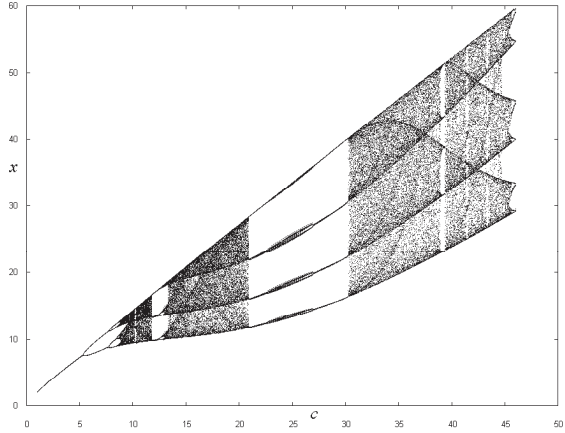}
\caption{\label{fig:bif}Bifurcation diagram for the Rossler
attractor when $a=b=0.1$ and $c$ is varied.}
\end{figure}

In most chaotic schemes, though, the key space is nonlinear
because all the keys are not equally strong. We say that a key is
\emph{weak} or \emph{degenerated} if it is easier to break a
ciphertext encrypted with this key than breaking a ciphertext
encrypted with another key. There exist keys giving rise to
non-uniformly distributed chaotic values. As shown in
Fig.~\ref{fig:bif}, bifurcation diagrams help to discover the
intervals for which a given parameter originates periodic orbits.
These values should be avoided, and the chaotic bands should be
preferred.

When many parameters are used simultaneously as part of the key,
the mutual interdependence complicates the task of deciding which
are the good intervals. In any case, the authors of the proposed
cryptosystem should conduct a study of the chaotic regions of the
parameter space from which valid keys, i.e., parameter values
leading to chaotic behavior, can be chosen. Depending on the
number of parameters chosen as part of the key, this region will
have $1,2,3,\dots$ dimensions.

A possible way to describe the key space might be in terms of
positive Lyapunov exponents. According to \cite[p. 196]{caitds},
let $\mathbf{f}$ be a map of ${\mathbb{R}}^m$, $m\geq 1$, and
$\{{\mathbf{x}}_0,{\mathbf{x}}_1, {\mathbf{x}}_2,\dots\}$ be a
bounded orbit of $\mathbf{f}$. The orbit is chaotic if

\begin{enumerate}
    \item it is not asymptotically periodic,
    \item no Lyapunov exponent is exactly zero, and
    \item the largest Lyapunov exponent is positive.
\end{enumerate}

\begin{figure}[t]
\center \includegraphics{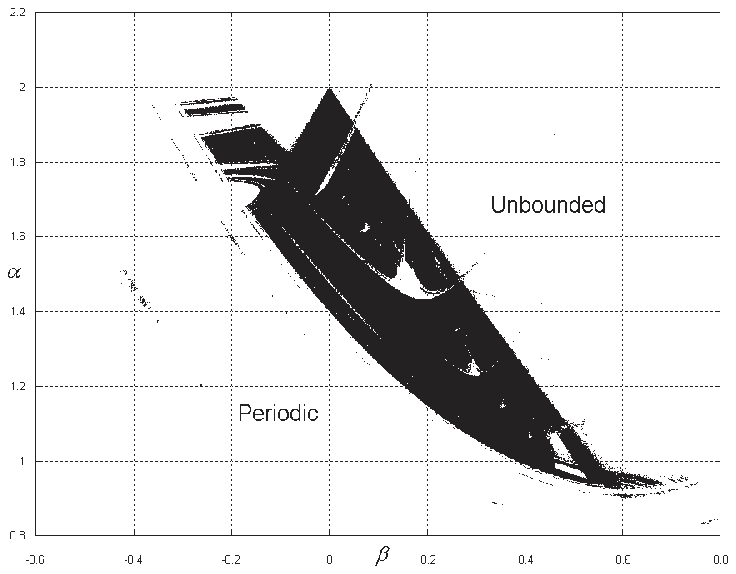} \caption{\label{fig:lyap}Chaotic
region for the Henon attractor.}
\end{figure}

The largest Lyapunov exponent can be computed for different
combinations of the parameters. If it is positive, then the
combination can be used as a valid key. In Fig.~\ref{fig:lyap},
the chaotic region for the Henon attractor has been plotted
following this criterion. This region corresponds to the keyspace.
In general, parameters chosen from the lower white region give
rise to periodic orbits, while parameters chosen from the upper
white region give rise to unbounded orbits. Both regions should be
avoided to get suitable keys. Only keys within the black region
are good. And even within this region, there exist periodic
windows, unsuitable for robust keys.

\begin{figure}[t]
\center \includegraphics[width=0.7\columnwidth]{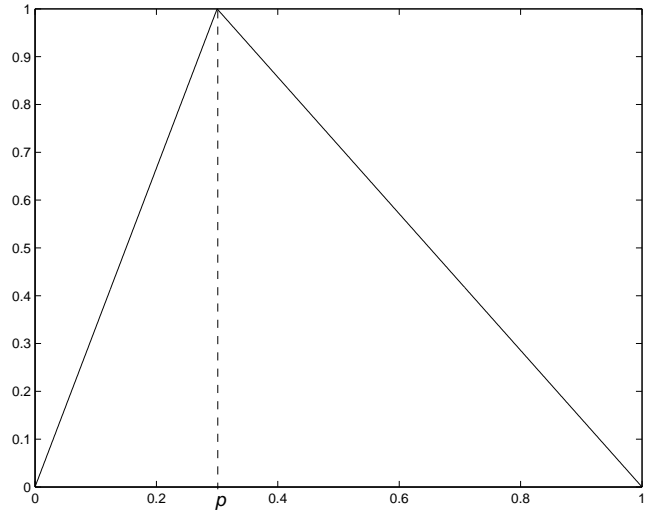}
\caption{\label{fig:skew_tent_map}Skew tent map with a control
parameter $p$.}
\end{figure}

However, this type of irregular and often fractal chaotic region
shared by most secure communication systems proposed is inadequate
for cryptographic purposes because there is no easy way to define
its boundary. Complete chaoticity for any parameter value should
be preferred. A simple map which behaves in this way is the
following skew tent map with a control parameter $p\in(0,1)$ (see
also Fig. \ref{fig:skew_tent_map}):
\begin{equation}
F(x)=
\begin{cases}
x/p, & x\in[0,p]\\
(1-x)/(1-p), & x\in(p,1]
\end{cases}.
\end{equation}

For any control parameter $p\in(0,1)$, the above piecewise linear
map is always chaotic. In fact, for a piecewise linear chaotic map
$F:X\to X$, if each linear segment is mapped onto $X$, the map
will be chaotic and have many desired dynamical properties
\cite[\S3.2.1]{ShujunLi:Dissertation2003}. Following such a
result, as long as the control parameter does not change the
mapping of each linear segment onto $X$, the obtained chaotic map
will be good to construct chaotic cryptosystems.

\begin{rules}
The chaotic region which constitutes the key space $\mathbf{V}$
from which valid keys are to be chosen should be thoroughly
specified.
\end{rules}

It is sometimes taken for granted that the security of the
encryption scheme is related to the size of the key space. A
necessary, but not sufficient, condition for an encryption scheme
to be secure is that the key space be large enough to frustrate
brute force attacks (see Sec.~\ref{sec:bfa}). If the chaotic
region does not meet this requirement, then it should be enlarged
accordingly. However, the solution is not as simple as
discretizing the region with a finer grain, because this could
lead to \emph{equivalent} keys, i.e., the same ciphertext is
decrypted by two different keys. When two different keys are very
close to each other, they could decrypt part or all the
ciphertext. The safeguard between adjacent keys should be defined.
In other words, the chaotic region should be discretized. In
chaotic regime, the sensitivity to parameters will guarantee that
two orbits starting from the same initial point but with slightly
different parameters will diverge (the divergence rate is given by
the Lyapunov exponents). However, the need for synchronization
sometimes allows for an important parameter mismatch.

From a cryptographic point of view, the secret parameter should be
sensitive enough to guarantee the so-called avalanche property:
even when the smallest change occurs for the parameter, the
ciphertext will change dramatically. Ideally, mathematical
expectation of the change of ciphertext is half of the maximal
value of all possible ciphertexts. For example, a natural idea to
do so is to iterate the employed chaotic system for multiple times
\cite{pwtcisea}.

\begin{rules}
The useful chaotic region (the key space $\mathbf{V}$) should be
discretized in such a way that the avalanche effect is guaranteed:
two ciphertexts encrypted by two different keys $v_1,v_2$ chosen
as close as possible from the key space $\mathbf{V}$ will be
entirely different.
\end{rules}

In some chaotic cryptosystems where more than one parameter is
used as part of the key, it is possible to fix one of them and try
to approximate the others. This is an undesirable behavior.
Ideally, when the key is made from a number of parameters, the
recovered signal after an illegal decryption attempt should never
reveal that the attacker is approaching the exact key when one
parameter is slowly varied. In other words, the recovered signal
should appear the same if none of the parameters is guessed as if
all but one of the parameters are guessed. This implies that the
total key space is the product, and not the addition, of all the
parameters involved.

\begin{rules}
Partial knowledge of the key should never imply partial knowledge
of the cleartext.
\end{rules}

\subsection{Key generation}

Once the key has been defined and the key space has been properly
characterized, the process to choose good keys should be explained
in detail. If some parameter ranges are given and the parameter
values can be randomly chosen within these ranges, then it should
be clear that there is no possibility of generating weak or
degenerated keys.

Sometimes the useful chaotic region has irregular shapes. This
shape could be enclosed in a regular one, such as a sphere or a
cube, and the key could be chosen randomly within this regular
shape and checked whether it is in the useful chaotic region.

\begin{rules}
The algorithm or process to generate valid keys from the key space
$\mathbf{V}$ should be thoroughly specified.
\end{rules}

\section{Security analysis}
\label{sec:analysis}

When a new cryptosystem is proposed, it should always be
accompanied by some security analysis. Although this analysis
cannot comprise all the possible attacks against the newly created
cipher, it should include at least the best known attacks and the
corresponding results. This analysis helps to spot and correct
flaws before the new scheme is published. That a cipher is
resilient to all these attacks does not mean that it is secure,
but at least it has undergone a certain amount of critical
analysis. It is a necessary, but not sufficient, condition for
security.

First of all, to resist common attacks, the designed cryptosystem
should have the following basic cryptographic properties:
confusion and diffusion. The first property reflects uniformity of
all keys, and the second one reflects strong sensitivity
(avalanche) of the key to small change. Obviously, Rule 6
mentioned above corresponds to diffusion. Here, we add a rule
corresponding to confusion. To achieve confusion, statistical
properties of the ciphertext, such as distribution, correlation
and differential probabilities, should be independent of the exact
value of the key. Many chaos-based cryptosystems are broken
because of the lack of confusion property, such as E. Alvarez et
al.'s cipher proposed in \cite{natce}.

\begin{rules}
For different keys, no distinguishable difference of the
ciphertext should be found from statistical point of view.
\end{rules}

Next we describe the sort of attacks which should be accounted
for.

\subsection{Classical types of attacks}

When cryptanalyzing an encryption algorithm, the general
assumption made is that the cryptanalist knows exactly the design
and working of the cryptosystem under study, i.e., he knows
everything about the cryptosystem except the secret key. This is
an evident requirement in today's secure communications networks,
usually referred to as Kerchoff's principle \cite[p. 24]{ctap}.
According to \cite[p. 25]{ctap}, it is possible to differentiate
between different levels of attacks on cryptosystems. They are
enumerated as follows, ordered from the hardest type of attack to
the easiest:

\begin{enumerate}
    \item Cipher text only: The opponent possesses a string of
    cipher text, $c$.
    \item Known plain text: The opponent possesses a string of
plain text, $p$, and the corresponding cipher text, $c$.
    \item Chosen plain text: The opponent has obtained temporary access to the
encryption machinery. Hence he can choose a plain text string,
$p$, and construct the corresponding cipher text string, $c$.
    \item Chosen cipher text: The opponent has obtained temporary access to the
decryption machinery. Hence he can choose a cipher text string,
$c$, and construct the corresponding plain text string, $p$.
\end{enumerate}

In each of these four attacks, the objective is to determine the
key that was used. The last two attacks, which might seem
unreasonable at first sight, are very common when the
cryptographic algorithm, whose key is fixed by the manufacturer
and unknown to the attacker, is embedded in a device which the
attacker can freely manipulate. Daily life examples of such
devices are electronic purse cards, GSM phone SIM (Subscriber
Identity Module) cards, POST (Point Of Sale Terminals) machines,
or web application session token encryption.

Many examples of how to break chaotic cryptosystems with known
plaintext and chosen plaintext attacks can be found in
\cite{coaces,kcoaccm,coacscs,coaecc,coadccuek}.

\begin{rules}
At a minimum, it should be checked whether the cryptosystem is
broken by simple known plaintext and chosen plaintext attacks.
\end{rules}

\subsection{Chaos-specific attacks}

Many different methods have been proposed to attack chaotic
encryption schemes, both for analog and digital systems. In this
paper we will focus on the former. There are three possibilities
for their cryptanalysis \cite{cocborcr}:

\begin{enumerate}
    \item The extraction of the message signal $m(t)$ from the transmitted ciphertext
signal $c(t)$.
    \item The extraction of the chaotic masking signal $x(t)$.
    \item The estimation of the parameters of the chaotic
    receiver, which are chosen from the key space $\mathbf{V}$.
\end{enumerate}

The extraction of the message signal is generally possible if
$m(t)$ is a periodic signal or consists of periodic frames with
sufficient duration. It can be accomplished using different
methods: spectral analysis techniques \cite{bcscuas},
autocorrelation and cross-correlation \cite{pwtcisea}, filtering,
or time series power estimation.

The chaotic masking signal $x(t)$ can be extracted using time
delaying embedding reconstruction or return maps
\cite{uamccs,stusc,emmbc,ccscurm}.

Parameter identification can be performed using generalized
synchronization \cite{bcsugse} or spectral analysis. Sometimes it
is not even necessary to use approximate parameter values because
transmitter and receiver are synchronized even under severe
parameter mismatch \cite{pwtcisea}.

\begin{rules}
At a minimum, it should be checked whether the cryptosystem is
broken by well known chaos-specific attacks.
\end{rules}

Also, as we mentioned above, when chaotic systems are implemented
in digital form, the dynamical degradation can also cause security
defects, which may bring weak keys and make brute force attacks
easier. For examples with security defects caused by dynamical
degradation of digital chaotic systems, please see
\cite{otsoacespwccifcp,LiShujun:JEIT2003} (or Chap. 4 of
\cite{ShujunLi:Dissertation2003}).

\subsection{Application-specific attacks}

In some specific applications, there are some special attacks. For
example, for digital images (videos), unlike normal
one-dimensional data, strong correlation always exists between
different pixels (transform coefficients). Such correlation
information can be used to develop some correlation-based attacks,
if the information is not successfully cancelled in
cipher-images/cipher-videos.

In addition, in some applications cryptosystems should be
specially optimized to make the encryption more efficient.
However, sometimes there exists a tradeoff between efficiency and
security, and the increment of efficiency causes decrement of
security. For example, encrypting partial data of plain-videos is
very useful to promote the encryption speed and make the
cryptosystem practical, but it is possible to reconstruct partial
visual information of plain-videos from partial non-encrypted
data.

It is true that different uses of a same cipher in different
applications may cause different levels of security
\cite{Schneier:Secrets&Lies2000}.

\begin{rules}
At a minimum, it should be checked whether the cryptosystem is
broken by well known application-specific attacks. If there exists
a tradeoff between efficiency and security, explicit criteria
should be given to show how to balance the two factors.
\end{rules}

\subsection{Brute force attacks}
\label{sec:bfa}

A brute force attack is the method of breaking a cipher by trying
every possible key. The quicker the brute force attack, the weaker
the cipher. Feasibility of brute force attacks depends on the key
space size $r$ of the cipher and on the amount of computational
power available to the attacker. Given today's computer speed, it
is generally agreed that a key space of size $r<2^{100}$ is
insecure.

\begin{rules}
To prevent brute force attacks, the key space size should be
$r>2^{100}$.
\end{rules}

However, this requirement might be very difficult to meet by some
proposed ciphers because the key space does not allow for such a
big number of different strong keys. For instance,
Fig.~\ref{fig:lyap} was created using a resolution of $10^{-3}$,
i.e., there are $1400\times 3000$ different points. To get a
number of keys  $r>2^{100}\simeq 10^{30}$, the resolution should
be $10^{-15}$. However, with that resolution, thousands of keys
would be equivalent, unless there is a strong sensitivity to
parameter mismatch, which is usually lost by synchronization.

\section{The channel} \label{sec:channel}

Many secure communications systems proposed are tested in Matlab
or some other simulation program, but not under real conditions.
It should be noted that a real channel is subjected to noise, has
a limited bandwith, and is attenuated. When designing a new secure
communications scheme, besides security considerations, the
channel characteristics should be taken into account, thus
preventing the system from failing when tested in a noisy,
bandwith-limited, and attenuated channel.

\begin{rules}
The secure communications system should work in a channel with
$-30$ dB signal/noise ratio, with limited bandwith, and with an
attenuation between $-3$ dB and $+1$ dB.
\end{rules}

\section{Conclusions}
\label{sec:conclusion}

We have presented a set of guidelines which are recommended to be
adopted by every secure communications systems designer. These
guidelines do not constrain the freedom and creativity of the
designer, but, if followed, guarantee a reasonable degree of
security and cryptographic rigor. In this way, new chaos-based
cryptosystems will be easier considered by the cryptography
community, and both worlds will be benefited.

\section*{Acknowledgements}

This research was supported by Ministerio de Ciencia y
Tecnolog\'{\i}a, Proyecto TIC2001-0586.

\end{document}